\documentclass[11]{article}

\usepackage{hyperref}	
\usepackage{amsmath}	
\usepackage{graphicx}	
\usepackage{fullpage}	
\graphicspath{{picture/}}
\usepackage{ amssymb }
\usepackage{feynmf}

\begin{document}

\title{Large space shell model calculations with small space results }

\author{Xiaofei Yu and Larry Zamick\\
 \textit{$^{1}$Department of Physics and Astronomy, Rutgers University,
Piscataway, New Jersey 08854, USA} }

\date{\today}
\maketitle

\begin{abstract}
We note that in large space shell model calculations and experiment
one sometimes get results , the form of which also appear in smaller
space calculations. On the other hand there are some results which
demand the large space approach.

\end{abstract}

\section{Introduction}
Let us start by saying that large shell model calculations are absolutely
essential for getting quantitative results for comparison with experiment.
And indeed model spaces are getting larger and larger.Three body interactions
are being included and ab into calculations are being performed. If
one examines the wave functions of these large space calculations one
finds very little of the so called leading configurations -the ones
that were used in the early days. One might expect that by this time
the whole shell model would be destroyed yet somehow this is not the
case, and this is what we will here address this problem.

\section{Quadrupole moments and magnetic moments}

We start with experiments and calculations that were performed for
the semi-magic N=28 isotones by Speidel et al. \cite{Speidel}. The focus
was on magnetic moments of excited 2+ states of even-even nuclei but
the ground stated even=odd were also included in the analysis. There
is a theorem that in the single j shell of particles of one kind all
g factors are the same. we are here dealing with close shell of neutrons
and valence protons in the $f_{7/2}$ shell. In the simple Schmidt
model the values of the $g$ factors would be $(3+2.793)/3.5= +1.655$.
However as seen in ref\cite{Speidel} the overall values are less than this
and are not constant. It was long ago pointed out by Arima , Horie
and Noya \cite{Arima1}\cite{Arima2}\cite{Arima3} in a first order perturbation calculation that
the $g$ factors would be quenched. Not only that but the quenched would
be n dependant. The more protons in the $f_{7/2}$ shell the more
quenching. The experimental values for both the even=even and even odd
nuclei show this behaviour. Furthermore large space calculations
beyond first order perturbation theory, although giving quantitatively
somewhat different values for the $g$ factors , more or less preserve
the overall picture of g factors lying on a straight line with a negative
slope. In T. Ohtsubo et al.\cite{Ohtsubo} the missing $\mathrm{^{49}Sc}$ moment
is measured with a value of $5.615(35)$. A theoretical analysis which
is more or less identical to the one made previously by Speidel et
al.\cite{Speidel} was also performed for the even- odd nuclei only, however
with theoretically obtained renormalized magnetic moment operators,

As another example we discuss quadrupole moments of ground states
of even odd nuclei in the Calcium isotopes. In a recent publication
R.F. Garcia\cite{Ruiz} presented in part measurements of this for
$\mathrm{^{43}Ca}$,$\mathrm{^{45}Ca}$ and $\mathrm{^{47}Ca}$ (also ontside the $f_{7/2}$
shell $\mathrm{^{49}}$Ca and $\mathrm{^{51}Ca}$). In their Fig 4 they show that results.
The ground state quadrupole moments appear to lie roughly on a straight
line starting from negative at$\mathrm{^{41}Ca}$ to positive beyond $\mathrm{^{44}
Ca}$. We here point out that there is a well known formula which gives
this behaviour.
\begin{equation}
Q= - (2j+1-2n)/(2j+2){*} <r^{2}> e_{eff}
\end{equation}

This appears in Lawson's book \cite{Lawson} and was used by Robinson et
al. in a study of the Ge isotopes \cite{Robinson}. The results are also consistent
with the statement that in a single j shell of particles of one kind
$Q(hole)=-Q(particle)$. The $NN+3N$ calculations in the Garcia et al .\cite{Ruiz} paper
follow this trend although in detail there are some deviations. The
$Q$ values they give for $A=43$, $45$ ,and $47$ are $-0.0246$, $+0.0252$ and $+0.0856$.

When they come to $^{49}$Ca in the single j shell we have one $p_{3/2}$
particle so we expect a negative $Q$ whils for $\mathrm{^{51}Ca}$ which is
a $p_{3/2}$ hole a positive $Q$.This obtained in both the experiment
and $NN+3N$ calculation\cite{Ruiz}.

In this paper they also discuss magnetic moments of the even odd ground
states . The magnetic moments in the $NN+3N$ calculation appear to lie
on a staight line reminiscent of the first order calculations of Arima
et al.\cite{Arima1}\cite{Arima2}\cite{Arima3}. The calculated values for $A=43$, $45$ and $47$ in the
$NN+3N$ calculation are $-1.56$, $-1.45$ and $-1.38$ respectively. The experiment
shows a somewhat flatter behavior.

We now come to cases where there are major disagreements between experiment
and the single $j$ shell models and where large space calculations are
an absolute necessity.These concern the magnetic moments of excited
states of the $\mathrm{Ca}$ isotopes, $s$ $J=2^{+}$ states. Whereas the $g$ factor
of an $f_{7/2}$ neutron is negative $g=-1.193/3.5= -0.547$, the measured
values for the $2^{+}$states in the work of Speidel et al. \cite{Zamick}
are positive. We have a gross violation of the single $j$ shell model.
in that work one put in by hand about a $50\%$ admixture of highly
deformed states with $g=Z/A$ which is about 0.5. To round things out
a bit a $50\%$ admixture of single $j$ with a $g$ factor of $-0.5$ and a deformed
mixture of $+0.5$ leads to an overlaa $g$ factor of zero. It remains a
challenge to see if the ab initio calculations can handle these highly
deformed admixtures. In a work of Taylor et al.\cite{Taylor} it is noted
that the g factor of $\mathrm{^{46}Ca}$ returns to negative although still
far from the Schmidt model. this is an indication that $\mathrm{^{48}Ca}$
is a better closed shell than $\mathrm{^{40} Ca}$. 

The extreme non- perturbative behaviours for the $2^{+}$ states in
$\mathrm{^{42}Ca}$ and $\mathrm{^{44}Ca}$ suggest that deformed admixtures could
be of importance also for the previously mentioned ground states of
the even-odd Ca isotopes. We suggest that the flat behaviour for the
even-odd $\mathrm{Ca}$ isotopes rather than the downslope shown for the $N=28$
isotones by Speidel et al.\cite{Speidel} could be due to these non perturbative
admixtures. As an example in $\mathrm{^{41}Ca}$ the Schmidt magnetic moment
is $-1.913 \mu_{N}$ but experiment is $-1.58\mu_{N}$. In second
order perturbation theory Mavromatis et al. gets close to this result
\cite{Mavromatis} . However the inclusion of meson exchange currents gets one
back to square zero, very close to the original Schmidt value. See
for example the review by I.A. Towner \cite{Towner}.One can get back to
the experimental result by including about $15\%$ of the highly deformed
admixture.

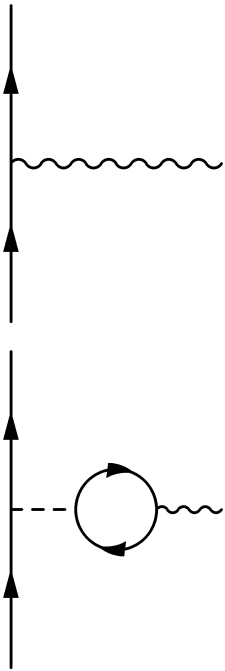
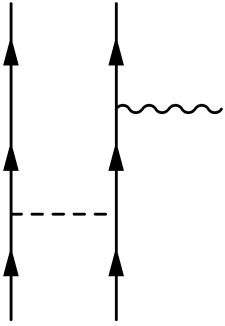
\begin{fmffile}{diagram}

\vspace{1em}

\begin{fmfgraph*}(80,120)
    \fmfstraight
    \fmfleft{i1,i2,i3}
    \fmfright{o1,o2,o3}
    \fmf{fermion}{i1,i2,i3}
    \fmf{photon}{i2,o2}
    \fmflabel{$X$}{o2}
\end{fmfgraph*}

\vspace{1em}

\begin{fmfgraph*}(80,120)
     \fmfstraight
    \fmfleft{i1,i2,i3}
    \fmfright{o1,o2,o3}
    \fmf{fermion}{i1,i2,i3}
    \fmffreeze
    \fmf{dashes}{i2,v1}
    \fmf{photon}{v2,o2}
     \fmf{fermion,left,tension=0.4}{v2,v1,v2}
      \fmflabel{$X$}{o2}
\fmf{phantom,tension=5}{v1,v1}

\end{fmfgraph*}

\vspace{1em}

\begin{fmfgraph*}(80,120)
  \fmftop{o1,o2,o3}
  \fmfbottom{i1,i2,i3}
  \fmf{fermion}{i1,v1,v2,o1}
  \fmf{fermion}{i2,v3,v4,o2}
  \fmffreeze
  \fmf{dashes}{v1,v3}
  \fmfright{i3,o4,o5,o3}
  \fmf{photon}{v4,o5}
  \fmflabel{$X$}{o5}
\end{fmfgraph*}

\end{fmffile}


\begin{thebibliography}{10}
 
\bibitem{Speidel}
K.-H. Speidel, R. Ernst, O. Kenn, J. Gerber, P. Maier-Komor, N. Benczer-Koller, G. Kumbartzki, L. Zamick, M. S. Fayache, and Y. Y. Sharon Phys. Rev. C 62, 031301(R) (2000)

\bibitem{Arima1}A. Arima and H. Horie, Prog. Theor. Phys. 11, 509
\textasciitilde{}1954!

\bibitem{Arima2}A. Arima and H. Horie, Prog. Theor. Phys. 12, 623
\textasciitilde{}1954!

\bibitem{Arima3}H. Noya, A. Arima, and H. Horie, Suppl. Prog. Theor.
Phys. 8, 33 \textasciitilde{}1958!.

\bibitem{Ohtsubo}T. Ohtsubo et al. PRL 109, 032504 (2012)

\bibitem{Ruiz}R.F. Garcia Ruiz et al. Phys. Rev. C 91, 041304(R)
(2015)

\bibitem{Lawson}R.D. Lawson, Theory of the Nuclear Shell Model,
Clarendon Press, Oxford (1980)

\bibitem{Robinson} S.J.Q. Robinson, L. Zamick, Y.Y. Sharon Journal-ref:
Phys.Rev.C83:027302 (2011)

\bibitem{Zamick}K.-H. Speidel, S. Schielke, O. Kenn, J. Leske,
D. Hohn, H. Hodde, J. Gerber, P. Maier-Komor, O. Zell, Y. Y. Sharon,
and L. Zamick Phys. Rev. C 68, 061302(R) (2003)

\bibitem{Taylor}M.J. Taylor et al. Phys. Lett. B \underline{605}
(2005) 265-272.

\bibitem{Mavromatis} H.A. Mavromatis, L. Zamick and G.E. Brown Nucl.
Phys. 80,545 (1966).

\bibitem{Towner} I.S. Towner Phys. Rep. 155, 263 (1987)\end{thebibliography}
\end{document}